\documentclass[aps,pre,amsmath,amssymb,prb,twocolumn]{revtex4-2}
\usepackage{amssymb,amsmath,bm}
\usepackage{graphicx}
\usepackage{enumerate}
\usepackage{color}
\usepackage{hyperref}
\usepackage{ragged2e}
\usepackage{blindtext}

\begin{document}

\title{Quantum Transport and Spectroscopy of Two-dimensional Perovskite/Graphene Interfaces}

\author{Yan Sun}
\affiliation{LPS, Universit\'e Paris-Saclay, CNRS, UMR 8502, F-91405 Orsay, France}
\author{C. Morice}
\affiliation{LPS, Universit\'e Paris-Saclay, CNRS, UMR 8502, F-91405 Orsay, France}
\author{D. Garrot}
\affiliation{Université Paris-Saclay, UVSQ, CNRS, GEMaC, 78000, Versailles, France}
\author{R. Weil}
\affiliation{LPS, Universit\'e Paris-Saclay, CNRS, UMR 8502, F-91405 Orsay, France}
\author{K. Watanabe}
\affiliation{Research Center for Electronic and Optical Materials, National Institute for Materials Science, 1-1 Namiki, Tsukuba, 305-0044, Japan}
\author{T. Taniguchi}
\affiliation{International Center for Materials Nanoarchitectonics, National Institute for Materials Science, 1-1 Namiki, Tsukuba, 305-0044, Japan}
\author{M. Monteverde}
\affiliation{LPS, Universit\'e Paris-Saclay, CNRS, UMR 8502, F-91405 Orsay, France}
\author{A.D. Chepelianskii}
\affiliation{LPS, Universit\'e Paris-Saclay, CNRS, UMR 8502, F-91405 Orsay, France}

\begin{abstract}
Quantum transport properties in molecularly thin perovskite/graphene heterostructure are experimentally investigated by Shubnikov-de Hass (SdH) oscillation and photo-resistance spectroscopy. We find an efficient charge transfer between the perovskite nanosheets and graphene, with a high hole concentration in graphene of up to $\rm \sim 2.8 \times 10^{13}\ cm^{-2}$. The perovskite layer also increases Fermi velocity lowering the effective mass of graphene from expected $\rm \sim 0.12\ m_e$ to $\rm \sim 0.08\ m_e$. Combining magneto-resistance and density functional theory calculations, we find that the carrier density in graphene significantly depends on the perovskite termination at the interface, affecting the charge transfer process and leading to a coexistence of regions with different doping. We also investigate the photo-response of the SdH oscillation under illumination. Using photo-resistance spectroscopy, we find evidence of photo-assisted transport across the perovskite layer between two graphene electrodes mediated by hot carriers in perovskite. Our results provide a picture to understand the transport behavior of 2D perovskite/graphene heterostructure and a reference for the controlled design of interfaces in perovskite optoelectronic devices.
\end{abstract}


\maketitle
Halide Perovskite is an attractive new class of promising semiconductors for optoelectronic devices \cite{jeong2020stable, fakharuddin2022perovskite}. To improve the power conversion efficiency and stability, the incorporation of graphene into perovskite-based optoelectronics has been exploited, in which graphene is introduced into transport layers or as electrodes to adjust the optical and electrical properties of perovskite devices  \cite{pescetelli2022integration, petridis2018renaissance, agresti2019two, taheri2018graphene,  razza2021laser,lee2015high,cheng2016van}. Graphene is a monolayer structure of carbon atoms with good optical transparency, excellent stability, high mobility, and outstanding thermal and electrical conductivity \cite{nair2008fine, dean2010boron,geim2007rise}. Moreover, as an ideal 2D system with a Dirac electronic spectrum, graphene offers a unique platform for discovering novel physics \cite{novoselov2004electric, novoselov2005two}. 

Ruddlesden- Popper perovskites (RPPs) have recently become appealing candidates in optoelectronic devices, thanks to their improved stability and higher photoluminescence quantum yield, and compositional and dimensional tunability \cite{shi2018two, leng2020bulk}. 
RPPs have a general formula $ \rm LA_{2} A_{n-1} B_{n} X_{3n+1}$, where LA and A represent the organic ammonium cations, B is a divalent cation, X is a halide anion, and n is the number of inorganic layers between two planes of LA cations. The large-size organic protective layers in the RPPs enable the natural formation of quantum well structure \cite{wang2016perovskite, leng2018molecularly}, the integration with other 2D materials  \cite{leng2018molecularly, leng2020electron, pan2021deterministic}, and promising wide-range applications. However, the interlayer charge transfer hindered by the organic barriers limits the performance of the devices. In this context, graphene has been introduced as a low-resistance contact to form 2D perovskite/graphene field effect transistors owing to the atomically smooth interface and better energy level alignment \cite{leng2020electron, shao2017stable, tan2016two, qiu2023interfacial}. Nevertheless, in graphene-integrated perovskite heterostructures, the microscopic mechanism of electronic properties and interfacial interactions are poorly understood, which obstructs the development of design principles of perovskite devices, partly due to the difficulty of building clear interfaces and keeping phase purity during processing because of the instability of perovskites with humidity and heat. In particular, the impact of the perovskite layer on the transport properties of graphene has not yet been investigated. 

Here we study quantum transport properties in heterostructure of molecularly thin perovskite $\rm (BA)_2(MA)_3Pb_4I_{13}\ ( BA= CH_3(CH_2)_3NH_3;\ MA= CH_3NH_3^+)$ and monolayer graphene by Shubnikov-de Hass (SdH) oscillation and photo-resistance spectroscopy. We report that the perovskite efficiently transfers charges to graphene, leading to a high doping level in graphene of up to $\rm 2.8 \times 10^{13}\ cm^{-2}$. This charge transfer also lowers the effective mass (cyclotron mass) of graphene from the expected theoretical effective mass of 0.12 $m_e$ to 0.08 $m_e$. The mobility in graphene is expected to be low for interfacial charge transfer doping because the charges in the perovskite can act as Coulomb scattering centers reducing mobility. However, we find that graphene exhibits a higher-than-expected carrier mobility of $\rm 550\ cm^{2} V^{-1} s^{-1}$. We performed density functional theory (DFT) simulations of the charge transfer between graphene and perovskite layer finding hole-dopped graphene allowing us to explain the doping levels observed in the experiment. These simulations show that the doping is not defect-mediated and that the potential in graphene created by perovskite electrons is periodic. This can explain the renormalized effective mass and observed mobility. The DFT simulations also show 
that the doping strongly depends on the microscopic structures of the interface.  While the interfaces with Pb-I and $\rm I^{-}$ termination lead to a high charge transfer, the charge transfer is much less efficient for the structure with BA molecules at the interface. Magneto-resistance and micro-fluorescence (micro-PL) experiments confirm the coexistence of different doping regions in the heterostructure, which is thus directly related to the interfacial terminations of the perovskite. 
Finally, we perform photo-resistance experiments at a high magnetic field in the regime of SdH oscillations related to the formation of Landau Levels. Under a laser excitation, we observe SdH oscillations in photoresistance which are phase-shifted from the resistance. This can be interpreted as a combined consequence of the photogating effect on the interfaces and a slight rise in graphene carrier temperature. We then perform spectroscopy experiments using a tunable wavelength excitation source allowing us to measure the photo-resistance spectrum. We show that the magnetic field dependence of this spectrum provides direct insight into the nature of the hot-carrier mediated photo-assisted transport in the graphene/perovskite/graphene device.

\section{graphene and 2D perovskite heterostructure.}

$\rm (BA)_2(MA)_3Pb_4I_{13}$ (BMPI) crystals with n=1,2,3,4 are prepared by temperature-controlled crystallization. The clean periodic diffraction peaks in the X-ray diffraction (XRD) pattern in Fig S1 confirm the phase purity of those 2D perovskite crystals. We used the crystal with n=4 which can be exfoliated even down to monolayer. Since generally perovskite is sensitive to heat and humidity, developing a method to avoid degradation during fabrication is crucial. To minimize heating during the assembly process, we used resist-free dry transfer, so that the transfer temperature remains lower than 60 $\rm ^ \circ C$. The photoluminescence spectrum of the bulk crystal and monolayer BMPI shows a single narrow peak at around 664 nm (Fig S1). 
To verify the quality of the BMPI layer after the exfoliation and dry-transfer process, we performed micro-PL microscopy on a pristine BN/graphene/BMPI/BN sample. The two layers of BN are used to encapsulate the whole stack. As shown in Fig S2, The mean photoluminescence intensity of BMPI/graphene is almost half as large as that of BMPI. A quenching of BMPI PL by graphene is expected since graphene has no band gap and is rendered as a collector for both electrons and holes, which is also proof of a clean interface between BMPI and graphene.\cite{niu2015controlled} The spatial fluctuations of the PL in the BMPI/graphene region are twice larger compared to BMPI. The characteristic domain size is around 2 $\rm{\mu m}$, we will discuss the possible origin of such domains later. All the above characterization results show that our crystals and fabrication technique, including exfoliation and dry transfer, ensure a good sample quality, making transport measurements possible.

\begin{figure*}[t!]
\centerline{
\includegraphics[clip=true,width=16cm]{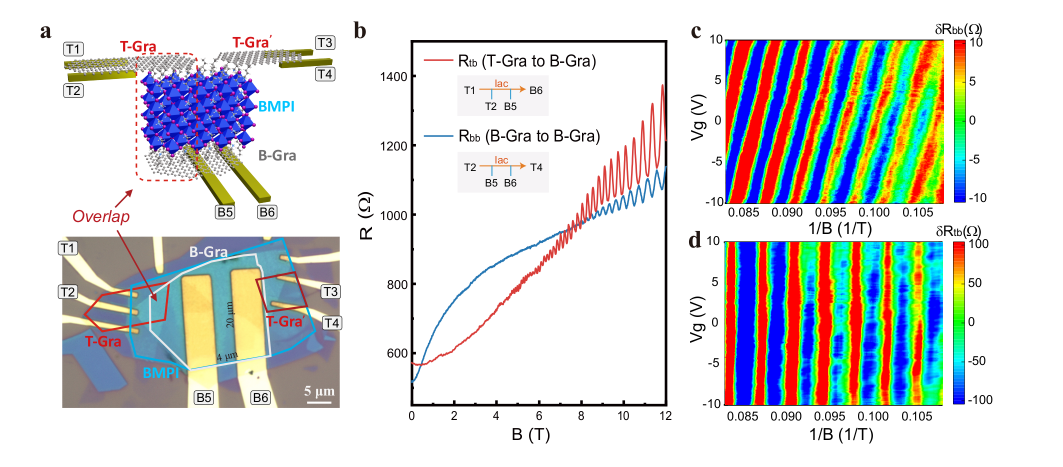} 
}
\caption{\justifying a) Schematic illustration (top) and micrograph (bottom) of a graphene/BMPI/graphene device, where a vertical region is constructed by introducing an overlap from top graphene (T-Gra) to bottom graphene (B-Gra). The whole device is encapsulated by two pieces of BN layers and stacked on a SiO$_2$/Si substrate, which also works as a bottom gate. The electrodes used in four-terminal connections are labeled. b) The magneto-resistance of BMPI/graphene stack at 1.6 K from B-Gra to B-Gra ($R_{bb}$) and T-Gra to B-Gra ($R_{tb}$) contacts with gate bias Vg = 0 V, both of which have the characteristic of the Shubnikov-de Hass (SdH) oscillations. The resistance of $R_{bb}$ is multiplied by 5 (the aspect ratio in device geometry). SdH oscillation as a function of gate bias, c) from $R_{bb}$, and d) from $R_{tb}$. In both cases, the graphene is heavily hole-doped with carrier density $n \rm \simeq 2.78 \times 10^{13}\ cm^{-2}$, deduced from the oscillation period. c) The SdH in $R_{bb}$ shift with bottom gate voltage $V_g$ indicating hole doping, d) while SdH oscillation in $R_{tb}$ shows a barely visible gate voltage dependence. This is most likely due to the screening of the gate voltage by the bottom graphene. } 
\label{fig:fig1}
\end{figure*}

A schematic and optical image of a graphene-BMPI device used in transport experiments is shown in Fig. \ref{fig:fig1}a. The device containing both horizontal and vertical stacking regions is constructed by transferring the BMPI flake (around 4 layers, thickness about 14 nm) on top of a graphene layer, followed by partly covering the stack with another small piece of graphene. Here an overlap region between top and bottom graphene through the BMPI layer is introduced. The stack is encapsulated between two h-boron nitride layers (Top-BN and Bottom-BN). The electrodes conducting the top and bottom graphene were fabricated on top of the bottom-BN layer before transferring graphene and BMPI, to avoid heating of the BMPI and contamination of the graphene surface during electrode fabrication. The bottom graphene electrodes are wider for better adhesion of the electrodes on the bottom BN layer and to ensure continuity across the bottom BN step.
The long aspect ratio of the bottom electrodes was chosen to have a well-defined geometry to measure bottom graphene conductivity with a large contact area between graphene and electrodes. For top graphene contacts that lay flat on the silicon surface, thinner electrodes were used and the geometry was adapted to the shape of the top graphene flakes.
The device is placed on a highly doped silicon wafer with 280 nm SiO$_2$, acting as a gate electrode. The details of materials and device fabrication can be found in the methods section. 

We investigate transport from top to bottom graphene through BMPI as well as the conductivity of the bottom graphene covered by the BMPI layer. Four-terminal electrical contacts are used to eliminate the contact resistance and resistance from wires. The resistance of the bottom graphene, $R_{bb}$, is normalized by the geometrical aspect ratio (5 squares). The resistance from top to bottom graphene, $R_{tb}$, is measured from the region T-Gra (top graphene) to B-Gra (bottom graphene), which has a larger overlap area rather than the region T-Gra'(another top graphene) to B-Gra (Fig.\ref{fig:fig1}a). The bottom graphene was measured at room temperature before deposition of the BMPI layers and exhibits a sharp Dirac peak and a  small electron doping (Fig S3). After stacking with BMPI, the Dirac peak strongly shifted, possibly indicating the charge transfer process. 
To confirm and quantify the charge transfer in the device, we cooled the sample to 1.6 K. The resistance of the samples decreased only weakly $\sim 15\%$ during cooling (Fig S3) for transport from top to bottom graphene. This indicates that the BMPI is sufficiently conducting even at low temperature and that transport through BMPI is not thermally activated. This also confirms that we prepared clean graphene/BMPI interfaces.
In the following, we present magneto-transport to elucidate the nature of the possible charge transfer process between graphene and BMPI, and photo-resistance spectroscopy to probe the interaction between photo-excitations in BMPI and transport in graphene.

\section{Magneto-transport at the graphene-perovskite interface}

We measured magneto-resistance and quantum oscillations to probe the carrier density in the graphene-perovskite heterostructure as well as its quantum transport properties. Fig. \ref{fig:fig1}b shows the resistance as a function of the perpendicular magnetic field, with four-terminal contacts on $R_{bb}$ (blue) and $R_{tb}$ (red). Below 1.6 K, the periodic oscillation on the magneto-resistance originates from the SdH effect, which is related to the crossing between the Fermi level and quantized Landau Levels (LL). The appearance of SdH oscillation in this hybrid graphene/perovskite system under an intermediate field ($<$8 Tesla) implies that the interface between graphene and BMPI is clean.

According to Lifshitz-Kosevich theory, the SdH oscillations in the magneto-resistance of two-dimensional electron gas (2DEG) systems can be described approximately by Eq.(\ref{eq1}) \cite{coleridge1989low, fujita2014theoryos}. In the limit of small SdH, their amplitude $\delta R$ is given by: 

\begin{equation}
\delta R = R_{0} D_T \exp \left(-\frac{\pi}{\mu B}\right) \cos\left( \pi\frac{h n}{2 e B} \right).\label{eq1}
\end{equation}

Here $n$ is the carrier density which fixes the SdH period, $\mu$ is the carrier mobility, and $R_{0}$ is the mean longitudinal resistance around which magneto-resistance oscillates. The resistance reaches a maximum when the number of occupied LL is a half-integer and the Fermi energy is at the center of a LL. The phase of the oscillation takes into account the Berry phase $\pi$ for graphene \cite{zhang2005experimental}. 
The temperature damping is described by the  Dingle factor $D_{T}$:
\begin{equation}
D_T = \frac{2 \pi^2 k_b T/\hbar \omega _c}{\sinh(2 \pi^2 k_b T/\hbar \omega _c)}  \label{eq2}
\end{equation}

where $\omega_c = e B/m_c^*$ is cyclotron frequency, and $m_c^* = \hbar k_F / v_F$ is the effective mass (cyclotron mass) \cite{neto2009electronic}. 

The resistance versus 1/B measured with gate voltage ranging from -10 V to 10 V displays a clear Landau fan diagram (Fig. \ref{fig:fig1}c). The explored gate range is rather narrow as we found that BMPI heterojunctions show irreversible changes in their transport properties even for moderate gate voltages of around 30 V, so we stayed far from this threshold in our experiments. 
Applying Eq.(\ref{eq1}), the oscillation pattern for $R_{bb}$  (Fig. \ref{fig:fig1}c) and $R_{tb}$ connections (Fig. \ref{fig:fig1}d) are simulated (Fig S4) and the carrier densities are extracted. We find that in both cases, the graphene is heavily hole-doped, with density $n$ $\rm \simeq 2.78 \times 10^{13}\ cm^{-2}$ at $Vg$ = 0 V. This doping level is so high that the hole density is about 1$\%$ of the areal density of carbon atoms in graphene (which is about $\rm \sim 3.8 \times 10^{15}\ cm^{-2}$).The typical distance between two holes is around 2 nm in the graphene, while the carbon-carbon bond length in graphene is around 0.14 nm, and the lattice constant of BMPI is around 0.87 nm {\cite{cooper2012experimental, shao2022unlocking}}. Considering the non-equivalent fabrication process for top and bottom graphenes, the nearly equivalent doping level in both layers suggests that the high hole density in graphene is independent of the fabrication process, but an intrinsic property of BMPI/graphene hybrids.  The peaks of SdH oscillations shift and the period slightly increases in $R_{bb}$ when increasing the gate voltage, as expected in quantitative agreement with the change in hole density (Supporting Information). For $R_{tb}$, the oscillation hardly depends on the bottom gate voltage. We attribute this to a dominating contribution of the top graphene/BMPI interface to the resistance, which is screened from the gate voltage by the bottom graphene. The different contribution between the top and bottom graphene can be attributed to the electrode geometry, as the resistance of the top and bottom graphene are expected to add since they are connected in series, but the geometrical factor for the two layers is different with a lower contribution from bottom graphene due to its wide bottom contact parallel to the overlap region.

\begin{figure}[t!]
\centerline{
\includegraphics[clip=true,width=8cm]{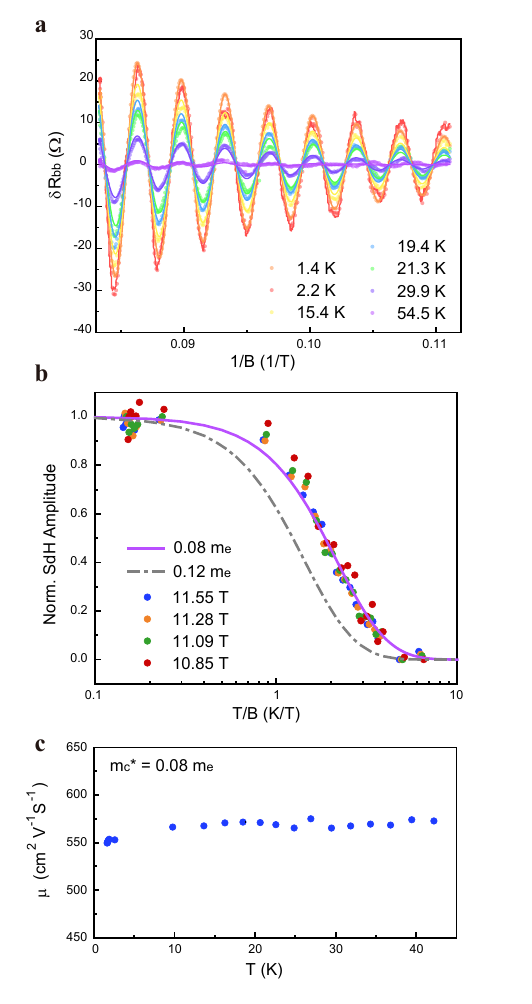} 
}
\caption{\justifying The effective mass($m_c^*$) and mobility extracted from SdH oscillation in magnetoresistance of $R_{bb}$. a) Temperature-dependent SdH oscillations which are fitted with Eq.(\ref{eq1}). b) The normalized amplitudes of the oscillations as a function of temperature/magnetic field are collected under several oscillation peaks labeled by the magnetic field, from which the $m_c^*$ is available through Eq.(\ref{eq2}). The curve with 0.08 $m_e$ is shown in a purple solid line. The broken-line curve shows the theoretical graphene effective mass for the sample hole density (with 0.12 $m_e$) as a comparison. c) With $m_c^* =0.08\ m_e$ the SdH mobility is independent of temperature (in agreement with $R(T)$ at $B=0$ T in this temperature range) with a value of around $\rm \sim 550\ cm^{2}\ V^{-1}\ s^{-1}$} 
\label{fig:fig3}
\end{figure}

The effective mass of monolayer graphene depends on the carrier density through $m_c^*= \hbar k_F$/$v_F$=$\hbar \sqrt{\pi n}$/$v_F$.\cite{neto2009electronic} Bring the carrier density $n$ = $\rm 2.78 \times 10^{13}\ cm^{-2}$ into calculation, the theoretical effective mass $m_c^*$ = 0.12 $m_e$, where $m_e$ is the electron mass, and the Fermi velocity $v_F$ is around $1 \rm \times 10^{6}\ m \ s^{-1}$ \cite{elias2011dirac}. One can also extract the effective mass from the amplitude pre-factor $D_T$ as a function of temperature. Fig. \ref{fig:fig3} shows that the amplitude of SdH oscillations, $\delta R_{bb}$, gradually decreases with increasing temperature, and tends to disappear under sufficiently high temperature ($>$ 54.5 K), where $\delta R_{bb}$ is calculated by subtracting the resistance with a liner fit to the background. The temperature-dependent SdH oscillations are fitted with Eq.(\ref{eq1}). To estimate the effective mass, we traced the amplitude of the oscillation peaks at various temperatures and normalized the amplitude according to the saturated resistance at the lowest temperature (Fig. \ref{fig:fig3}b). The only fitting parameter $m_c^*$ is available from the relation between the amplitude and T/B by Eq.(\ref{eq2}). For comparison, the curve corresponding to $m_c^*$ = 0.12 $m_e$ is shown in a broken line. We find that $m_c^*$ = 0.08 $m_e$ is the best fit for several oscillation peaks.

Further, we extracted the mobility from the exponential dependent in Eq.(\ref{eq1}). As shown in Fig S7a, in the case where $m_c^*$ = 0.12 $m_e$, the mobility increases by about 60 $\%$ with temperature rising from 1.6 K to 40 K, while, for $m_c^*$ = 0.08 $m_e$, the mobility remains almost unchanged with temperature( Fig. \ref{fig:fig3}c). In Fig S7b, the resistance also does not show significant monotonic changes with temperature from 1.6 K to 40 K, both with or without magnetic fields, which is consistent with the experimental effective mass $m_c^*$ = 0.08 $m_e$. The lower effective mass indicates an increase in the graphene Fermi-velocity to $1.5 v_F$. Enhanced Fermi velocities of up to $3 v_F$ have been reported previously as a result of enhanced electron-electron interactions near the Dirac point or reduced screening on substrates with a lower dielectric constant  \cite{hwang2012fermi,elias2011dirac}. In our sample interactions effects can instead enhanced through the interaction of holes with electrons in the BMPI layer which is also highly doped through charge transfer. 
The mobility $\mu$ at around 550 $\rm cm^{2}\ V^{-1}\ s^{-1}$ in our system corresponds to a mean free path of $40$ nm which is more than 10 times larger than the average distance between holes in graphene ($2\;{\rm nm}$) which 
is also the mean distance from electrons in BMPI. If the electrons in BMPI act as randomly distributed Coulomb scattering centers (all within the Fermi wavelength of the graphene plane, $\lambda_F \approx 6.7\rm \ nm$), the calculation of the expected mobility in our case following \cite{nomura2006quantum} gives mobility as low as $\rm 1\;cm^{2}\ V^{-1}\ s^{-1}$.
A spatially ordered charge distribution in BMPI near the interface, such as we find in DFT below, induces a periodic potential landscape explaining the reduced disorder compared to a random (defect-mediated) distribution of charges. This microscopic order in the BMPI potential structure can also contribute to the reduction of effective mass compared to its theoretical value.

\section{Interfacial terminations affect the charge transfer process}

\begin{figure}[h]

\includegraphics[clip=true,width=8cm]{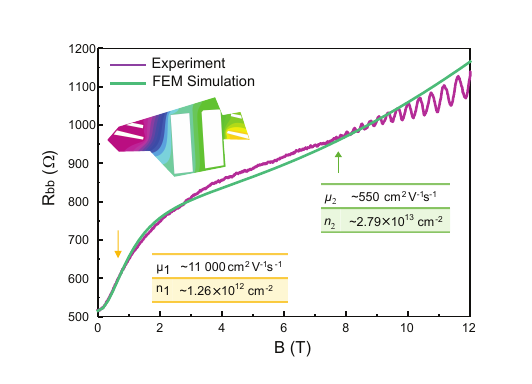}      
\caption{\justifying a) Simulations of magneto-resistance of $R_{bb}$ with a model with two carrier of different mobility. The yellow arrow point to the region where the higher mobility carriers contribute most to magnetoresistance while the green arrow point to a region of dominating contribution from lower mobility carriers. The mobilities and doping levels are shown in the tables. The simulated voltage drop on a model geometry from the finite element method (FEM) under 12 Tesla is shown in the inset as an example.
}
\label{fig:fig3v1}
\end{figure}

In addition to quantum oscillations, the magneto-resistance in Fig.\ref{fig:fig1}b shows a global positive magneto-resistance trend which is not expected in a single carrier Drude model. This behavior is specially marked in $R_{bb}$ but also visible in $R_{tb}$. The positive magneto-resistance hints that the carrier density in the sample may not be uniform, which can occur if different molecular structures of the BMPI-graphene interface co-exist in the device leading to different charge-transfer efficiencies. We thus decided to model the conductivity of our sample as the sum of two conductivities corresponding to the weakly/highly doped regions with densities $n_1/n_2$ and respective mobilities $\mu_1/\mu_2$. Naturally, we expect that the mobility in the weakly doped region will be higher ($\mu_1 > \mu_2$) due to reduced scattering on the charges in BMPI.
\begin{equation}
\sigma = n_1 e \mu_1 \frac{P}{1+i \mu_1 B} +  n_2 e \mu_2 \frac{1-P}{1+i \mu_2 B}
\label{eq:2mobilities}
\end{equation}
The expression Eq.~\ref{eq:2mobilities} gives the average mobility on length scales larger than the typical domain size which is around a $\mu m$ from micro-PL experiments. The spacing between the closest electrodes in our sample is around $5 \mu m$ which justifies this simplification.  
We fit the coverage fraction P of the two regions to the magneto-resistance and obtain P=0.4. To account for the complex sample geometry, we modeled the potential distribution in the sample by finite element method (FEM), calculating expected magneto-resistance for a conductivity given by Eq.~\ref{eq:2mobilities}. In this model, $n_2, \mu_2$ were fixed to their values obtained from the analysis of SdH oscillations while $n_1, \mu_1, P$ were used as fitting parameters.
We find that this two-region model can indeed reproduce the overall magneto-resistance trend (Fig. \ref{fig:fig3v1}). The high-doped region gives the saturating high field mobility, while the low-doping region of $\rm n_1 \approx 1.26 \times 10 ^{12} \; cm^{-2}$  with higher mobility of $\rm \mu _1 \approx 11000 \; cm^{-2}V^{-1}s{-1}$ explains the low field magnetoresistance. The estimated density in the low-doped regions is probably an average value, as it is very non-uniform due to strong density gradients from nearby high-density regions. This may also explain why we don't see any signature of SdH oscillations from the low-density regions despite their higher mobility.
Evidence of the co-existence of regions with different carrier densities is also found in the micro-PL microscopy on the pristine BN/graphene/BMPI/BN sample without electrodes, where we see inhomogeneous photoluminescence with domain size around 2 $\rm \mu m^2$ in BMPI/graphene heterostructure (Fig S2). 

A possible explanation for the coexistence of regions with different charge transfer densities is that the terminations of the perovskite layer contacting graphene could influence the charge transfer processes and change the doping level in graphene. Specifically, in the interface between graphene and perovskite, the termination of BMPI could be an organic BA molecular, inorganic Pb-I, or Idiode (structures shown in Fig S8). 
We test this hypothesis by calculating the charge density in graphene for different graphene/BMPI interfaces using density functional theory (DFT).

\begin{figure*}

\includegraphics[clip=true,width=16cm]{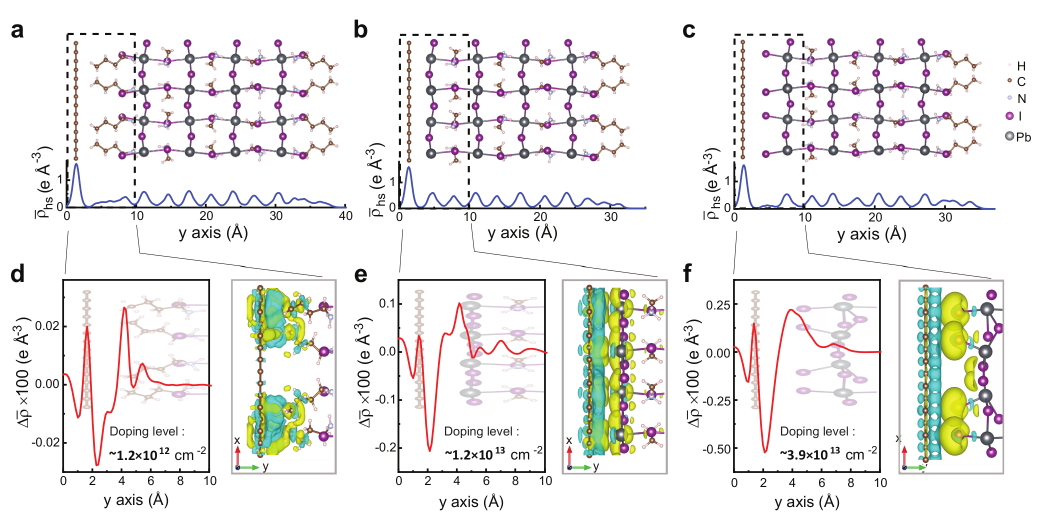}      
\caption{\justifying Density functional theory simulation of the electron density at the interface between BMPI and graphene for different perovskite terminations. a) BA molecules, b) Pb-I, c) Iodide ions. The blue curves denote the plane-averaged integral charge density along the stacking direction (y-axis) in the graphene/BMPI heterostructure ($\bar \rho_{\rm hs}$) with each type of interface. The charge transfer process mainly happens at the interface and is revealed by calculating the plane-averaged differential charge density ($\Delta  \bar {\rho}$). Positive (negative) values in $\Delta \bar{\rho}$ indicate charge accumulation (depletion). In d), e), and f), the $\Delta \bar{\rho}$ (red line, multiplied by 100) zoomed at the interface shows that the electrons are lost on the graphene side while accumulating on the BMPI side.  With BA molecular as the termination (d), a slight charge transfer corresponds to moderate hole doping in graphene. Whereas in the case of the Pb-I section (e) or I ions (f) as termination, a drastic transfer happens on the interface and results in an order of magnitude higher doping in graphene. The doping level of graphene in each case is labeled. The image (right-hand side) shows the atomic structure of the interface for each case and the color-coded density contours of $\Delta \rho$ from the x-y plane, the yellow (cyan) regions correspond to $\Delta \rho > 0 $ ($\Delta \rho < 0 )$.  
}
\label{fig:dft}
\end{figure*}

In our simulations, we consider microscopic structures with different types of termination of perovskite: BA (Fig.\ref{fig:dft}a), Pb-I (Fig.\ref{fig:dft}b), and I (Fig.\ref{fig:dft}c). Computational details can be found in the Experimental Section and Supporting Information. Because of the charge transfer process, the charge density in the heterostructure ($\rho_{\rm hs}$) is not equal to the charge densities for non-interacting graphene($\rho_{\rm graphene}$) and BMPI($\rho_{\rm BMPI}$). To show the charge carriers distribution, a differential charge density $\Delta  \rho$ is calculated, which is obtained from $\Delta \rho= \rho_{\rm hs} - \rho_{\rm graphene} - \rho_{\rm BMPI}$. We introduce the plane-averaged differential charge density $\Delta  \bar{\rho}$, defined as the stacking plane-average of $\Delta  \rho$, to estimate the magnitude of charge transfer,
\begin{equation}
\Delta  \bar{\rho}= \frac{1}{S_g}\int \Delta \rho \; dx dz ,
\label{eq5}
\end{equation}
The y coordinate is the stacking direction and $x, z$ are the coordinates within the graphene plane (the unit cell is shown in Fig.~\ref{fig:figs8}). The integral runs over slices of the unit cell for a fixed $y$, and $S_g$ is the graphene surface within the unit cell which is used for normalization. The plane-averaged integral charge density of the graphene/BMPI heterostructure ($\bar \rho_{\rm hs}$) is shown along the y-axis (blue line), which corresponds very well to each crystal structure. 

The charge transfer primarily arises on the interface between BMPI and graphene. This is illustrated in Fig. \ref{fig:dft}d,e,f, which shows that $\Delta \rho$ is concentrated near the graphene/BMPI interface. Positive values of $\Delta \bar{\rho}$ indicate electron accumulation, while the negative values represent electron depletion. 
For all the simulated interfaces, electrons are transferred from the graphene side to BMPI, leading to hole-doped graphene. The amplitude of this doping strongly depends on the type of termination.
The density contours on the right-hand side also reflect the differential charge density distribution in the x-y plane zoomed on the interface.

In Fig. \ref{fig:dft}d, only a small value of $\Delta  \bar { \rho}$ can be seen in the charge transfer from the BA molecular termination. This mild charge transfer corresponds to a doping level of around $\rm 1.2 \times 10^{12}\; cm^{-2}$ in graphene (see Supporting Information Section \ref{supinfoC} for details), which is close to the average density in the low doping region as estimated from magnetoresistance in Fig.\ref{fig:fig3v1}. In contrast, with the Pb-I termination (Fig. \ref{fig:dft}e), the doping level increases by an order of magnitude, reaching up to $\rm 1.2 \times 10^{13}\; cm^{-3}$. For this termination, the separation of charge accumulation and depletion layers becomes visible from the density contours of $\Delta \rho$. Finally in the case of $I^-$ termination (Fig. \ref{fig:dft}f), a considerable amount of charges accumulate around $I^-$, and results in a doping of $\rm 3.9 \times 10^{13} \;cm^{-2}$ in the graphene layer. We note that the density observed from the Shubnikov-de Haas oscillations is between these two high doping values. In the experiment, only a single SdH oscillation period is visible, which suggests only one type of high doping interface being realized in the experiment with a low enough disorder to exhibit quantum oscillations.


\begin{figure}
\includegraphics[clip=true,width=8cm]{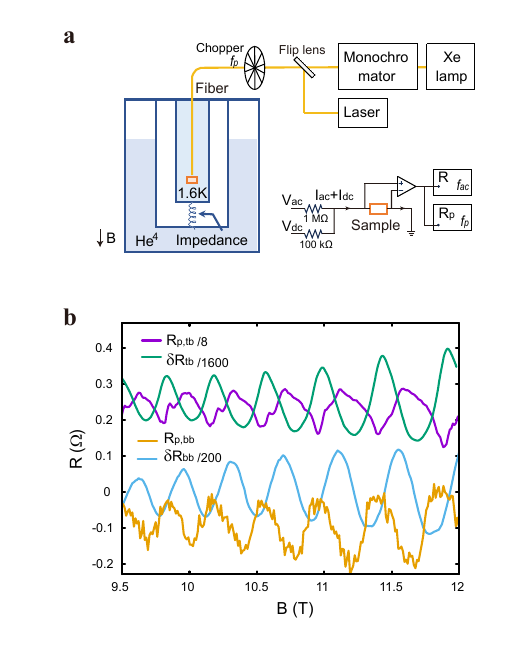}  
\caption{\justifying a)Schematic of the setup for the photo-excitation transport measurement, in which the illumination is laser or a Xenon lamp followed by a monochromator. The sample is immersed into superfluid helium 4 to accelerate thermal dissipation. The photoresistance is synchronized at chopper frequency ($f_p$) and biased by DC current. b) Photoresistance at high magnetic field in SdH region under excitaion of 532 nm laser with the power of 40 $\rm W m^{-2}$, for both B-Gra to B-Gra ($R_{p,bb}$) and T-Gra to B-Gra (R$_{p,tb}$), compared to the corresponding amplitude of SdH oscillation  $\delta R_{bb}$ and $\delta R_{tb}$. The $\delta R_{tb}$ and $R_{p,tb}$ are shifted for clarity. The $R_{p,tb}$ shows more asymmetrical oscillations, while $R_{p,bb}$ with weak phase shift from  $\delta R_{bb}$ due to the combination of photo-gating and heating.}
\label{fig:pv532}
\end{figure}

\begin{figure*}
\includegraphics[clip=true,width=16cm]{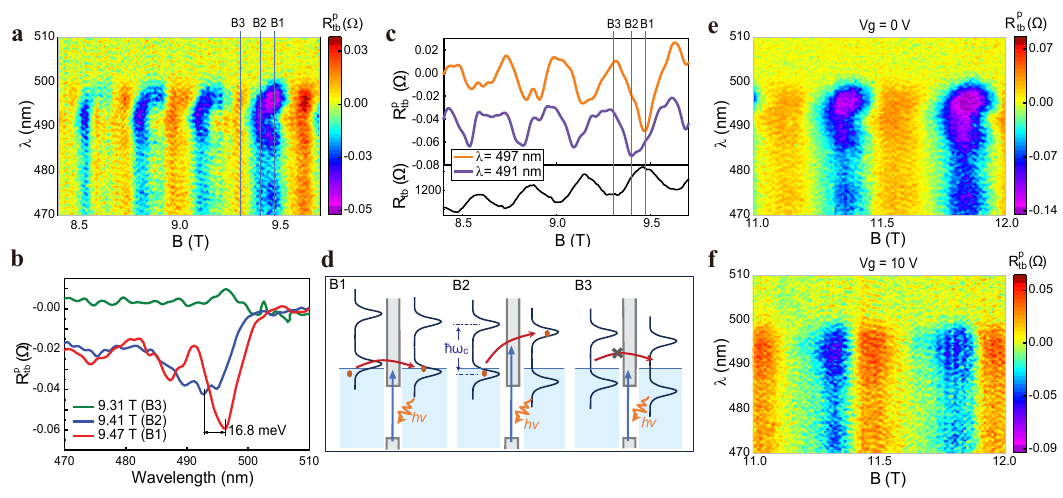}  
\caption{\justifying a) The change in resistance under illumination (R$^p_{tb}$) from $R_{tb}$ is plotted versus excitation wavelength on the vertical axis and magnetic field B on the horizontal axis. The numbers 1,2,3 label magnetic fields of 9.47 T ($B_1$), 9.41 T ($B_2$), and  9.31 T ($B_3$) respectively in one period.  b) Photo-resistance spectra show a shift of the absorption peak appearing from $B_1$ to $B_2$. The shift wavelength corresponds to an energy of 16 meV. c) Magnetic field dependence of the R$^p_{tb}$ show an enhanced amplitude at the maximum of R$_{tb}$ (the minimum of R$^p_{tb}$) compared to the amplitude expected from harmonic oscillations. d) A sketch of a qualitative model for photo-resistance. For $B_1$ the Fermi energy lies inside the LL for both top and bottom graphenes, this gives maximal photoassisted transport. For $B_2$ the Fermi energy enters the gap between LL in one of the graphene layers,   photo-assisted transport now requires an extra LL spacing energy $\hbar \omega_c$ to reach the first empty LL. This extra energy blue shifts the photo-resistance spectrum at $B_2$ compared to $B_1$ and the observed shift of $16\;{\rm meV}$ is indeed comparable to $\hbar \omega_c \simeq \;{\rm 14\;meV}$. Finally, for $B_3$, the Fermi energy lies in the gap between LL for both graphenes leading to a suppressed photo-resistance. Panels e) and f) show the comparison of photo-resistance peaks under $V_g = \;{\rm 0\; V}$ and f) $V_g = \;{\rm 10\; V}$. The spectral shift disappears with gate voltage. Since both BMPI and the top graphene are screened from $V_g$ by the bottom graphene, the gate voltage only changes the carrier density in the bottom graphene demonstrating that near alignment of the LL between top to bottom graphene is indeed required.}
\label{fig:rpmap}
\end{figure*}

\section{Photo-resistance spectrum on SdH oscillation}

For highly doped graphene the mean spacing between holes is around 2 nm, this is comparable to the exciton Bohr radius which is estimated to be about 1 nm in 2D RPPs and 4.2 nm in 3D perovskites \cite{blancon2018scaling, miyata2015direct}, primarily depending on permittivity and effective mass \cite{konzelmann2019quantum}. This implies the possibility of a strong interaction between carriers in graphene and excitons hosted by BMPI. 
To study photo-physics directly at the BMPI-graphene interfaces we performed photo-resistance measurements using a home-built cryogenic probe in which the sample was immersed in superfluid helium to ensure good thermalization under photo-excitation down to 1.6 K.
A sketch of the experimental setup is shown in Fig.\ref{fig:pv532}a. The sample was biased with both AC and DC currents, and two lock-in amplifiers were used to measure sample resistance at the same time($R$, from AC current and with frequency $f_{ac}$) and photoresistance at the chopper frequency ($R^p$, from dc current and with $f_{p}$). Since the heterojunctions are highly doped, the photo response was always a small change in the total sample resistance. 

We started by illuminating the sample with a 532 nm laser. The strongest photoresistance response appeared at high magnetic field in the SdH region. The photoresistance responses $R^p_{bb}$ and $R^p_{tb}$ are shown in Fig.\ref{fig:pv532}b together with the sample resistance $\delta R_{bb}$ and $\delta R_{tb}$, in which we see photo-resistance oscillations in the SdH region. 
From Eq.~(\ref{eq1})and (\ref{eq2}),  these can be attributed to a change in the carrier density (photo-gating) and the heating of graphene carriers. Heating leads to a decrease in the SdH amplitude but does not change the phase of the photo-resistance signal with respect to SdH ($R^p$ minima will be located at $R$ maxima). Photo-gating leads to a small change in the SdH oscillation period looking like the derivative of the SdH signal. 
The reduction (respectively increase) of the carrier density under illumination gives a photo-resistance contribution with a phase shifted by $\pi/2$ (respectively $3 \pi/2$). In the experiment, we find that $R^p_{bb}$ shows a $0.7 \pi$ phase shift with respect to $\delta R_{bb}$, which is consistent with a decrease in carrier density under illumination and some photo-induced heating. The out-of-phase component of photo-resistance (with respect to SdH) allows us to estimate the relative photo-induced change in the carrier density to be $\Delta n/n \simeq 2.5\times 10^{-5}$ (Supporting Information Section \ref{supinfoD}). For our laser excitation power of around $\rm 40\; W m^{-2}$, and assuming that only 15\% of the light is absorbed in the BMPI layer \cite{leng2018molecularly}, this gives a typical lifetime $\tau$ for electrons transferred from BMPI to graphene of around $\rm 400\;ns$ (this is probably a lower bound assuming that all the photons absorbed in BMPI contribute to photo-gating).
The in-phase photo-resistance gives a carrier temperature increase of $\rm 0.3\ K$ which is consistent with the heating expected from our laser excitation intensity (details of the estimates are given in supporting information).

In contrast to $R^p_{bb}$, the photo-response of the top-bottom graphene $R^p_{tb}$ is not simply a phase-shifted SdH signal: $R^p_{tb}$ oscillates very an-harmonic with pronounced negative peaks and some signatures of beating between two periods, which is not visible in $\delta R_{tb}$ (Fig.\ref{fig:pv532}). This difference can be related to the difference between photo-current (extraction of photo-excited electron-hole pairs from the BMPI) and photo-assisted transport (transfer of carriers between the two graphenes through excitation of BMPI) processes, which are allowed in two graphene-contacts geometry, but are absent in the single bottom-graphene geometry. Our sample is symmetric between top and bottom graphene and the applied DC voltages between top/bottom graphene are small (1 mV), thus strong photo-current response is not expected,  leaving photo-assisted transport as the most likely explanation for the strongly an-harmonic photo-resistance. 

To further understand the mechanism of photo-assisted transport in the heterostructure, we replaced the laser with a broadband white light source (Xenon lamp) combined with a monochromator, that allowed us to change the excitation wavelength continuously. The aspect ratio of $R_{bb}$ and the smaller SdH amplitude compared to $R_{tb}$ makes the detection of $R^p_{bb}$ more difficult, as seen in the larger noise in Fig.~\ref{fig:pv532}. The excitation power available after the monochromator (spectral excitation linewidth of 5 nm) was about 100 times lower than that from the 532 nm laser. Thus we focused on $R^p_{tb}$ for the photoresistance spectroscopy experiments. We note that in control experiments with a pristine graphene device, no measurable photoresistance spectrum is found under the same conditions.

The experimental results are shown in Fig.\ref{fig:rpmap}a as a function of both excitation wavelength $\lambda$ and magnetic field B.
At B1 = 9.47 T, a photo-resistance peak is observed around $497 \;{\rm nm}$, with a weak photo-resistance tail at longer wavelengths. The signal at $532\;{\rm nm}$ is no longer visible because of the much weaker excitation intensity from the monochromated Xenon lamp.
As the magnetic field increases to B2 = 9.41 T, a red-shift of the main absorption peak becomes visible at maxima of SdH for $R_{tb}$, disappearing rapidly away from the maximum. In Fig.\ref{fig:rpmap}b we show that this shift in photoresistance spectra appears at field $B_1 = 9.47\;{\rm T}$, but disappears already for $B_2 = 9.41\;{\rm T}$. The change in the magnetic field $B_1 - B_2 = 0.06\;{\rm T}$ is only 18\% of the total SdH period $0.34\;{\rm T}$ in this region of magnetic field. The shift in wavelength of the photo-resistance spectrum corresponds to an energy of around $\rm \Delta E = 16\;meV$, with the spacing between LL of around $\rm 14\; meV$, and a change of LL energy of around $\hbar e (B_1 - B_2)/m_c^* = 0.09 \rm\; meV$ from field $B_1$ to $B_2$. The spectral shift $\Delta E$ is thus much closer to the total LL spacing, suggesting that it does not come from a single graphene layer but instead from photo-induced transport between the top and bottom graphenes. 

A sketch of our qualitative model for photo-resistance is shown in Fig.\ref{fig:rpmap}d. When no gate voltage is applied the density of the top and bottom graphene are almost identical. Thus at SdH maxima, there can be direct photo-assisted transport between the two not completely filled LL (process 1).
The photon energy is then completely absorbed within the BMPI layer enabling charge transfer between the two graphenes. In this situation the magnetic field dependence of the $R^p_{tb}$ starts to show an enhanced amplitude at SdH maxima compared to the amplitude expected from harmonic SdH oscillations (see $R^p_{tb}$ at 497 nm at Fig.\ref{fig:rpmap}c), leading to the more anharmonic magnetic field dependence which was also seen with 532 nm laser excitation. 
As the magnetic field is changed slightly the LL in one of the two graphenes goes below the Fermi energy and an extra energy must be provided to transfer carriers to the first empty LL, requiring an increase in photon energy by the LL spacing $\hbar \omega_c$ (process 2). Finally, when LL on both graphenes are full at SdH minima the photoassisted transport is suppressed leading to weak photoresistance (process 3).

To check that the LL alignment between the two graphenes is important, we compared the signal at $V_g = 0\;{\rm V}$ (Fig.\ref{fig:rpmap}e) and $V_g =\;{\rm 10V}$ (Fig. \ref{fig:rpmap}f), leading to a different position of the Fermi energy with respect to LL in both graphenes. The gate voltage mainly changes the density of the bottom graphene and is screened in the rest of the device. Thus it mainly affects the relative position of LL between bottom and top graphene. Once the gate voltage is applied, the shift of the spectrum with the magnetic field disappears completely. Magnetic-field-dependent screening of Coulomb interactions with the BMPI layer by carriers in graphene can also lead to changes in the BMPI photo-resistance spectrum with the periodicity of SdH (high carrier density in graphene is expected to amplify such effects). In the current sample, these effects are difficult to distinguish given the stronger photo-assisted transport response.

The anharmonic photo-resistance SdH oscillatons and the energy associated with the spectral shift close to $\hbar \omega_c$ confirm our qualitative model for photo-assisted transport relying on the alignment of the LL between the two graphenes. This model implicitly assumes hot-carrier transport in BMPI, as the relative energy of the LL in both graphenes would not matter if carriers have to hop through many sites with random energies on the way. This may explain the blue-shifted onset for the main photo-resistance spectrum starting at around $500\;{\rm nm}$ (\rm$\sim 2.48\; eV$) compared to the expected position of the optical absorption spectrum for $n=4$ BMPI which is reported with bandgap of $\rm 2.07 \;eV$ {\cite{leng2020electron}}. The extra energy would then correspond to the excitation of higher-energy hot carriers in the BMPI layer. In this sample the LL are broad and their width is at the limit of our spectral resolution. However, we showed that this device has important density fluctuations and inhomogeneities. It is thus likely that, in future devices, individual LL will become more clearly visible, enabling spectroscopic investigations of the interplay between photo-excitation and transport between LL in the two graphenes. 

To summarize, we find that photo-resistance provides detailed information on the physics of photoexcitations at the graphene-BMPI interface, allowing us to estimate the characteristic lifetimes for photo-gating and heating of the graphene carriers. We also showed that photo-resistance spectroscopy provides evidence of direct photo-assisted transport between the two graphene contacts through the BMPI layer, possibly indicating that we excite hot carriers in BMPI that can interact with both top and bottom graphene.

\section{Conclusions}

We prepared nanosheet perovskite/graphene heterojunctions avoiding thermal degradation effects on the sensitive perovskite during fabrication. This allowed us to observe SdH oscillations in low-temperature magneto-transport, revealing a strong charge transfer process at BMPI/graphene interface, leading to a strong hole doping in graphene of up to $\rm\sim 2.8 \times 10^{13}\ cm^{-2}$. We find that the hole effective mass  0.08 $m_e$ is lower than its expected theoretical value of 0.12 $m_e$ for graphene at this density. The charge transfer also induces a high density of electrons in the perovskite, which is expected to substantially reduce the mobility in graphene if they act as random Coulomb scattering centers. However, we find higher mobility than expected $\rm \sim 550\ cm^{2} V^{-1} s^{-1}$, implying a potential ordering of the negatively charged scattering centers in the perovskite. This correlates with the change in effective mass which could be renormalized by the interaction with perovskite electrons. 
We also show that the charge transfer significantly depends on the termination of the perovskite at its interface with graphene. Density functional theory simulations show that different interfacial terminations lead to very different doping levels, in agreement with our magneto-resistance data and micro-fluorescence microscopy. 
The photo-resistance at high magnetic field in the regime of SdH oscillations is investigated, where the graphene LL become quantized. A phase shift appears on the photo-resistance under laser excitation compared with the sample SdH resistance oscillations, as the combined effects of the photo-gating effect and slight rise of carrier temperature. From the spectroscopy, the onset of photo-resistance shift by an energy similar to the LL spacing in graphene, suggests elastic (energy conserving) photo-assisted transport between the two graphene contacts through hot-carriers in BMPI.

Our experiments show that this system enables the exploration of new regimes in photo-assisted transport, and a possible access to new fundamental physics as the interface homogeneity improves in future devices.
Also, the strong built-in electric field associated with the efficient charge transfer can facilitate the separation of electrons and holes at the graphene/BMPI interface. This combined with the high hole mobility in graphene gives the hope of highly efficient nanosheet perovskite/graphene opto-electronic devices.

\section*{Methods}

{\bf Synthesis of BMPI crystals.}  
The synthesis of the BMPI single crystals was adapted from {\cite{leng2018molecularly}}. Generally, 154mg of PbO (0.69 M), 34mg of BAI (0.17 M), and 82 mg of MAI (0.52 M) precursors were mixed in a vial with 0.884 ml HI and 0.116 ml $\rm H_3PO_2$ solvent. The PbO can be substituted by the same molar amount of $\rm PbI_2$. A clear yellow solution is obtained after heating and stirring at $110\; ^\circ C$ for one hour. The stock solution was slowly cooled to room temperature at a rate of $\rm 1\; ^\circ C$/hour in an oven. The precipitated black crystals were vacuum filtered and rinsed with toluene several times, followed by drying in vacuum overnight.

{\bf Fabrication of Graphene/BMPI/Graphene device.}  
Au(30 nm)/Ti(5 nm) electrodes were deposited on a thin layer of flat hexagonal boron nitride (bottom BN), which had been transferred on a doped silicon substrate with a 300 nm-thick oxide layer. Then a monolayer graphene flake (bottom graphene) was transferred on top of electrodes and BN, and annealed under $\rm 200\; ^\circ C$ at $1 \times 10^{-5}$  mbar for 2 hours. Polydimethylsiloxane (PDMS) is used during mechanical exfoliation of BMPI flake (Fig S1). A BMPI flake on the PDMS surface was transferred to fully cover the bottom graphene. During resist-free dry-transfer, $\rm 60\; ^\circ C$ heating was used for a short time (around 1 minute) to help release the BMPI pieces. Afterwards, two top-graphene flakes were picked up at room temperature with a big top BN flake supported by PDMS. When stamping the top graphene/BN layer, we introduced a small overlap region between the top and the bottom graphene through BMPI layer. The exfoliation and dry-transfer processes were implemented in the air.

{\bf Transport measurements.}
Transport measurements are performed using SR830 lock-in amplifiers at an AC bias current of 100 nA, and a low-noise amplifier (LI-75A) with 100x Gain. The frequency for magnetoresistance is 17 Hz ($f_{ac}$ in Fig 5a), and for photo-resistance is 39 Hz ($f_p$) synchronized with the chopper frequency. Low-pass $\pi$ filters were added to each measuring line to protect the sample from noise and abrupt voltage discharges during connection switching.   
Photo-resistance measurements are performed in a pumped superfluid He4 chamber (temperature around 1.6 K). Helium flux into the chamber is fixed by an impedance with a room-temperature flow rate of around $3.5\times 10^{-2}\;{\rm mbar\; L \;s^{-1}}$.

{\bf First-principles computational methods.}
The charge density calculations in this work are based on density functional theory (DFT) conducted with plane-wave basis set and projector augmented wave (PAW) pseudopotentials implemented in the Quantum Espresso (QE) \cite{giannozzi2009quantum, giannozzi2017advanced}. The exchange-correlation function is taken as the Perdew–Burke–Ernzerhof (PBE) type of the generalized gradient approximation(GGA) \cite{perdew1996generalized}. DFT-d correction is used for all calculations to describe the van der Waals force between BMPI and graphene \cite{barone2009role}. To simulate the two-dimensional system and eliminate the interaction effects between the adjacent layers, the calculations were implemented with a vacuum thickness of 25 \AA. The cell parameters and atomic positions are fully relaxed with the k-point mesh of 3 $\times $ 1 $\times $ 6, generated by using the Monk-horst–Pack scheme. A kinetic energy cutoff of 550 eV is adopted for plane-wave expansion, and the convergence tolerances for the energy and force are set at 10$^{-8}$ eV. We used a Gaussian smearing with a deviation of 0.01 Ha. All the parameters were tested for convergence.

\bibliography{pegra.bib}

\vspace {5mm}

\noindent {\bf Acknowledgments}

The work was financially supported by funding from ANR-20-CE92-0041 (MARS), IDF- DIM SIRTEQ, and the European Research Council (ERC) under the European Union's Horizon 2020 research and innovation programme (grant Ballistop agreement no. 833350). We thank S. Gu\'eron, Z.Y.Chen, and E. Delporte for the discussions on the physical process. We also thank M. Entin for the help on modeling. 

\vspace {3mm}

\noindent{\bf Author contributions}

Y.S., M.M., and A.C. designed the experiments and implemented the magneto-transport measurements. Y.S. fabricated the devices used for the study and performed calculations of electron charge density under the guidance of C.M. D.G. carried out micro-PL measurements. R.W contributes to the technical assistance. K.W. and T.T. provide boron nitride crystals used in the experiment. A.C. and Y.S. exerted the Finite element method. All authors discussed the
results and contributed to the writing of the paper.


\clearpage 
\newpage
\onecolumngrid

\section{Supporting Information}

\subsection{Material information}

\begin{figure}[h!]
\centerline{
\includegraphics[clip=true,width=16cm]{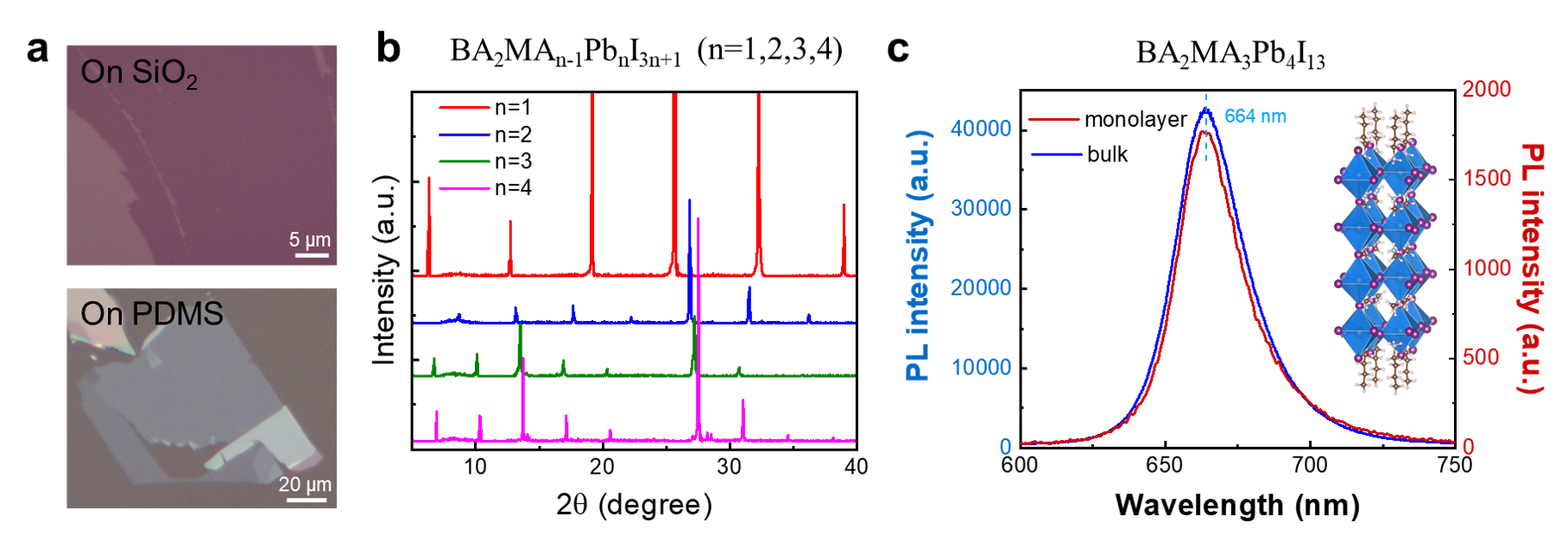} 
}
\caption{  \justifying  a) Examples of optical images of atomically thin BMPI flakes mechanically exfoliated on $\rm SiO_2$ substrate and PDMS substrate from bulk $\rm (BA)_{2}(MA)_{n - 1}Pb_n I_{3n + 1}$ single crystals. b) The X-ray diffraction pattern of $\rm (BA)_{2}(MA)_{n - 1}Pb_n I_{3n + 1}$ single crystals with n from 1 to 4 shows the phase purity of the crystals. c) Photoluminescence of $\rm (BA)_2(MA)_3Pb_4I_{13}$ shows single narrow peak  at $664 \; nm$ for both bulk crystals and exfoliated monolayer, in which the FWHM of monolayer BMPI is around $2 \; nm $ narrower 
. } 
\label{fig:figs1}
\end{figure}

\begin{figure} [h!]
\centerline{
\includegraphics[clip=true,width=16cm]{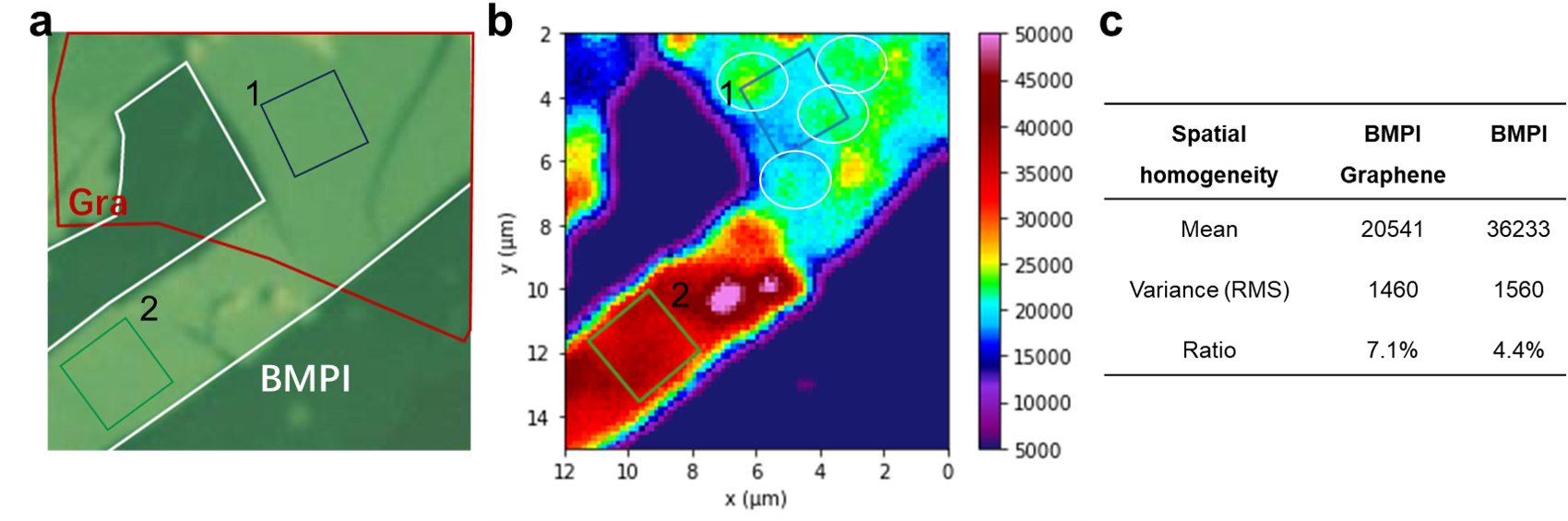} 
}
\caption{\justifying 
a) Optical image of a BN/graphene/BMPI/BN stack (BMPI on top of graphene), in which the two pieces of BN are utilized to encapsulate the whole stack. b) The corresponding micro-photoluminescence (micro-PL) map of the same region as a). The PL intensity in the BMPI-only region is almost twice as strong as that of graphene/BMPI as a result of a strong charge transfer process quenching fluorescence in the BMPI layer by graphene. Some island-distributed domains with stronger emission intensity are highlighted by white circles, and their typical size is around $ 2 \; \mu m^{2} $. To quantitatively compare those two regions, we selected the frame of the BMPI-only region (box 2 in green) and graphene/BMPI region (box 1 in dark blue). We extracted the variance by root-mean-square (RMS) and mean value of the fluorescence intensity, as shown in c). The homogeneity of  BMPI and BMPI/graphene can be evaluated by the RMS ratio $\rm (PL_{RMS}/ P_{mean})$ of the fluorescence intensity, which shows relatively homogeneous emission from BMPI layer $\rm (about\; 4.4\;\%)$ but less homogeneous emission from BMPI/graphene $\rm (about \;7.1\;\%)$. The inhomogeneous fluorescence in the graphene/BMPI stack could be related to the charge transfer process dominated by the type of perovskite termination at the interface. A $ 445\; nm$ ps laser was used as excitation with the power of $ 60\; \mu W cm^{-2}$ per pulse, frequency of 80 MHz, and integral time of 5 seconds. The micro-map was taken under a 4 K cryostat (Attocube attoDRY1000).
 } 
\label{fig:figs2}
\end{figure}

\begin{figure} [h!]
\centerline{
\includegraphics[clip=true,width=16cm]{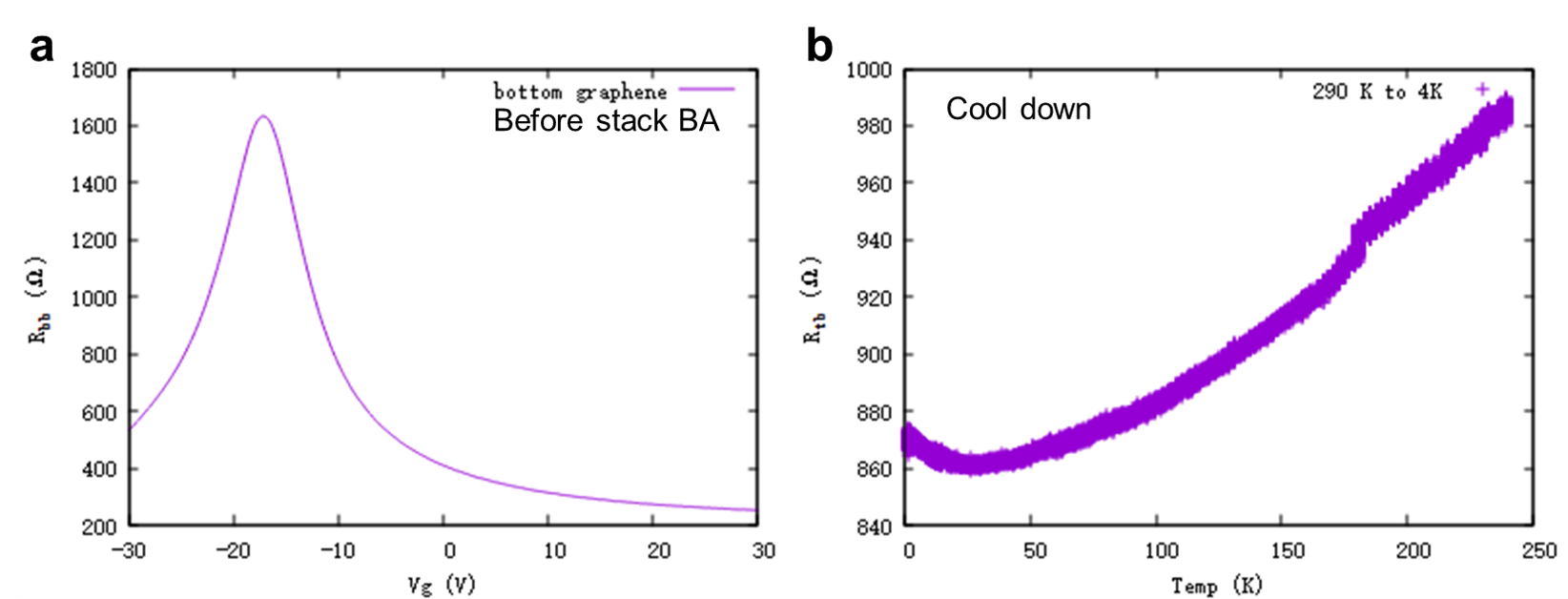} 
}
\caption{\justifying 
a) The transport curve of bottom graphene before putting BMPI under room temperature. The Dirac peak appears at around $\rm -17 \;V$, corresponding to electron doping of around $\rm 1.2 \times 10^{12}\; cm ^{-2}$. After depositing BMPI, the Dirac peak vanished indicating a strong charge transfer process. b) The temperature dependence of resistance from  T-Gra to B-Gra ($R_{tb}$). The weak changes during cooling indicate that the BMPI is sufficiently conductive even at low temperatures and that transport through BMPI is not thermally activated.
 } 
\label{fig:figs3}
\end{figure}

\newpage
\subsection{Measurements and analysis}

\begin{figure} [h!]
\centerline{
\includegraphics[clip=true,width=16cm]{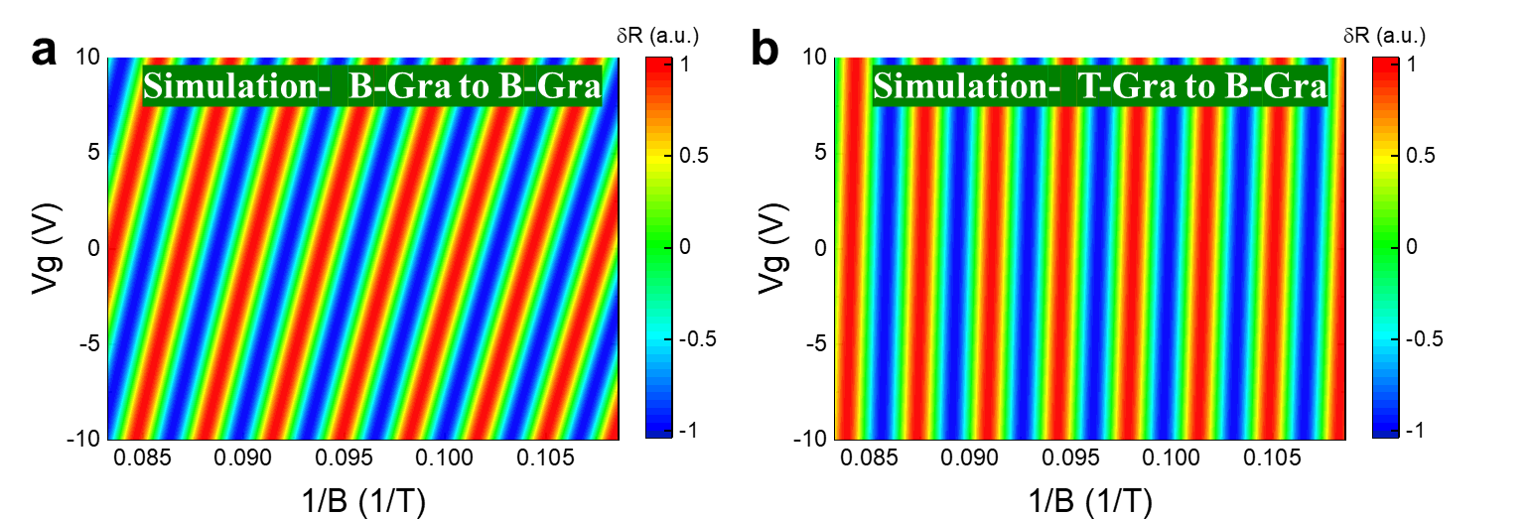} 
}
\caption{  \justifying 
Simulated SdH oscillation pattern versus gate bias with carrier density $ 2.8 \times 10^{13} \; cm^{-2}$ for a) B-Gra to B-Gra graphene ($R_{bb}$) and T-Gra to B-Gra graphene ($R_{tb}$) by applying the periodic function in Eq.(1), which is highly consistent with experimental data in Figure 1c and 1d. In the case of Rtb, the screen effect from the bottom graphene is considered and reflected on the smaller effective carrier density tuned by gate bias. 
 } 
\label{fig:figs4}
\end{figure}

\begin{figure} [h!]
\centerline{
\includegraphics[clip=true,width=16cm]{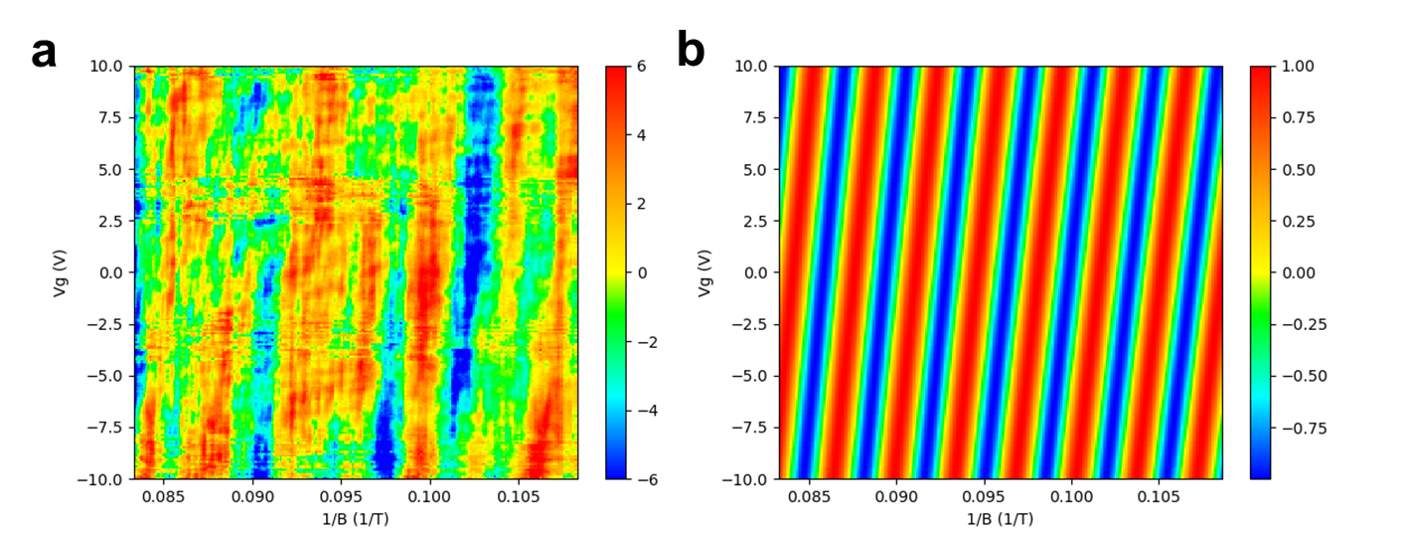} 
}
\caption{  \justifying 
a) A similar behavior can be seen on the junction from T-Gra' to B-Gra even if the SdH oscillations are less visible, due to the mesoscopic fluctuations in this sample with a smaller overlap and more disorder from gold electrodes. b) Simulated SdH oscillation pattern versus gate bias with carrier density $ 2.8 \times 10^{13} \; cm^{-2}$ and less screen effect considered.
 } 
\label{fig:figs5}
\end{figure}

\begin{figure} [h!]
\centerline{
\includegraphics[clip=true,width=16cm]{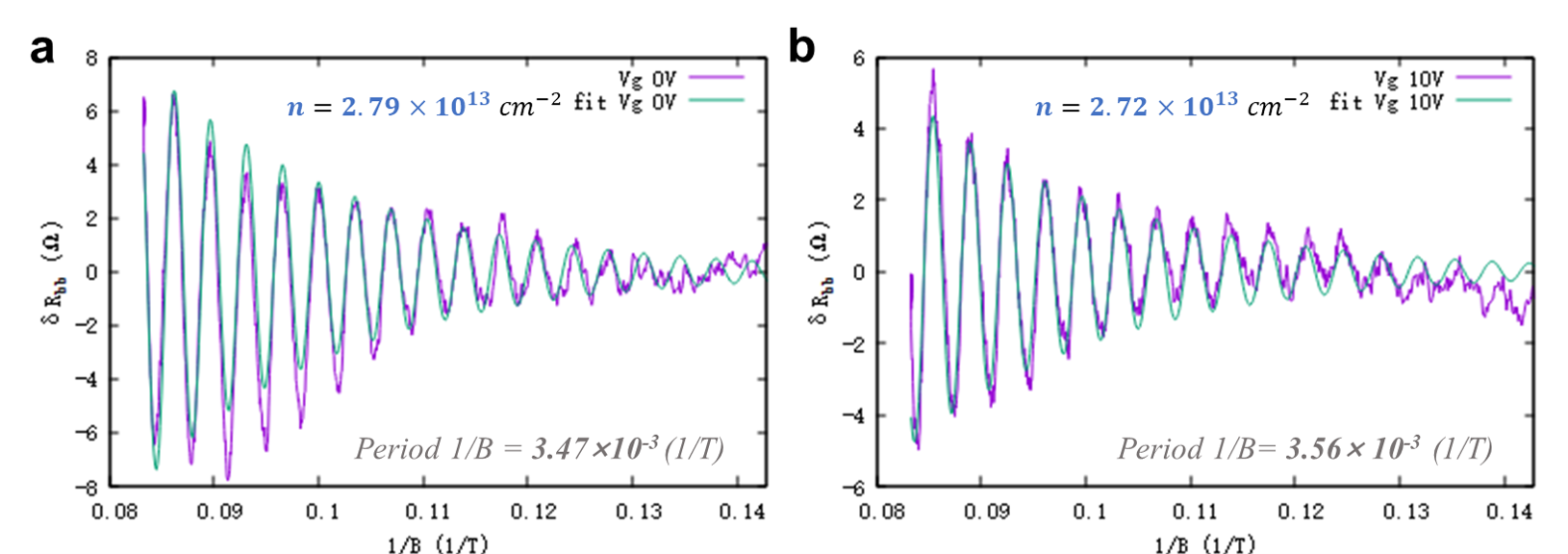} 
}
\caption{  \justifying 
Extraction of carrier density by fitting the SdH oscillation with Eq.(1) to obtain oscillation periods and carrier density under gate bias a) Vg = 0 V, b) Vg = 10 V.
 } 
\label{fig:figs6}
\end{figure}

\begin{figure} [h!]
\centerline{
\includegraphics[clip=true,width=16cm]{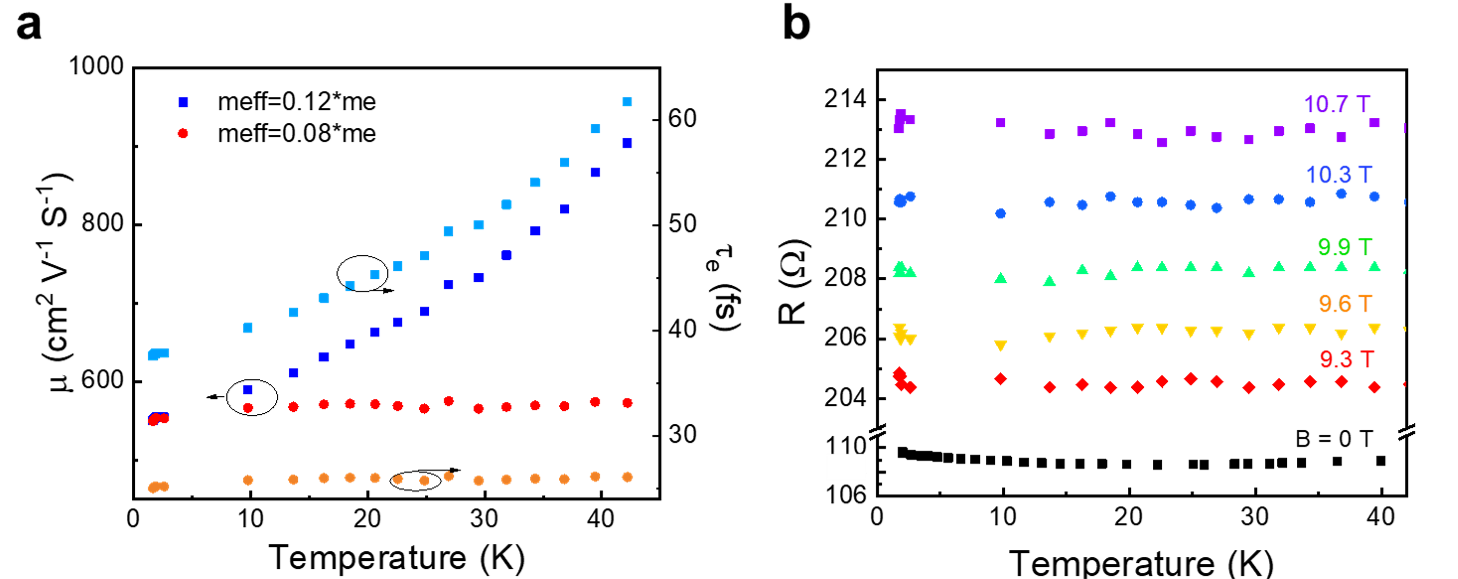} 
}
\caption{  \justifying 
a) The mobility keeps stable with temperature rising when $m* = 0.08\; m_e$, as well as the scatter lifetime $\tau_e$, while increases by about $60\%$ with temperature if $m* = 0.12\; m_e$. b) The temperature-dependent $R_{bb}$ under different magnetic fields shows no significant changes, which agrees with the experimental effective mass $m* = 0.08\; m_e$.
 } 
\label{fig:figs7}
\end{figure}

\newpage

\subsection{Theoretical results}
\label{supinfoC}

\noindent {\bf A.	Structural Models and Computational Details}

\vspace{2.5mm}

{\bf Commensurate structure }

\begin{figure} [h!]
\centerline{
\includegraphics[clip=true,width=8cm]{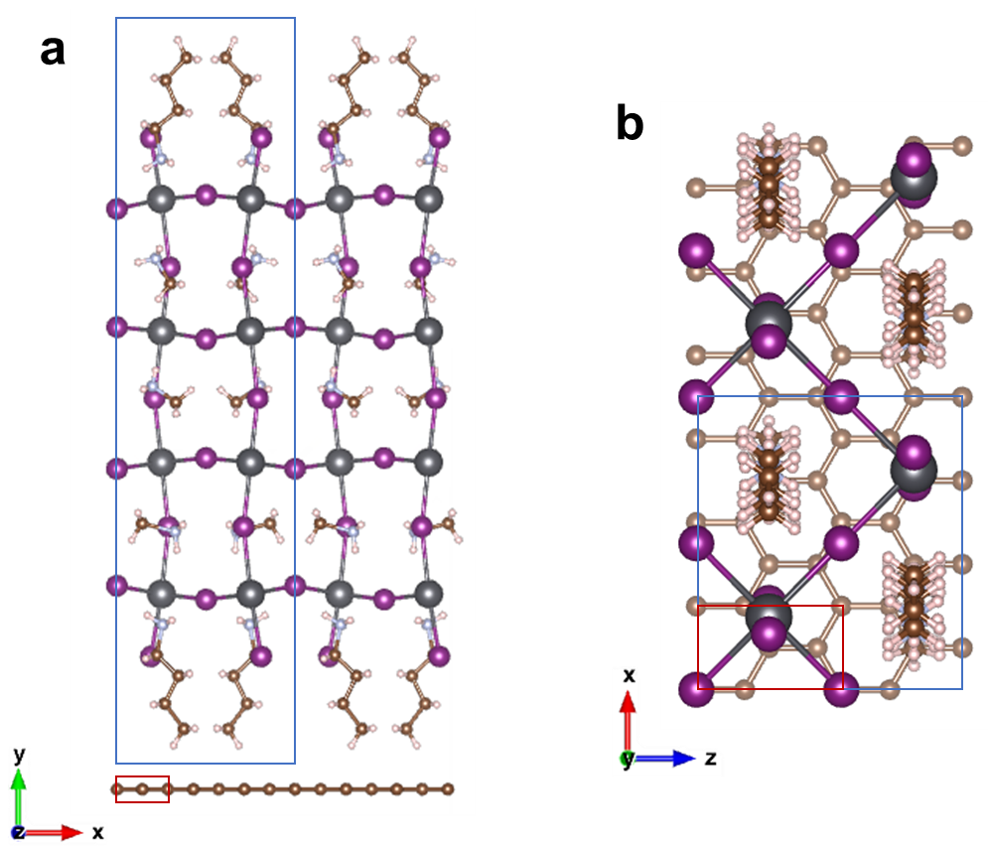} 
}
\caption{  \justifying 
Structural models of one cell of BMPI/Graphene heterostructures: a) side view and b) top view with BA termination. The used unit cells are indicated in red (for graphene) and blue (for BMPI).
 } 
\label{fig:figs8}
\end{figure}

\begin{figure} [h!]
\centerline{
\includegraphics[clip=true,width=14cm]{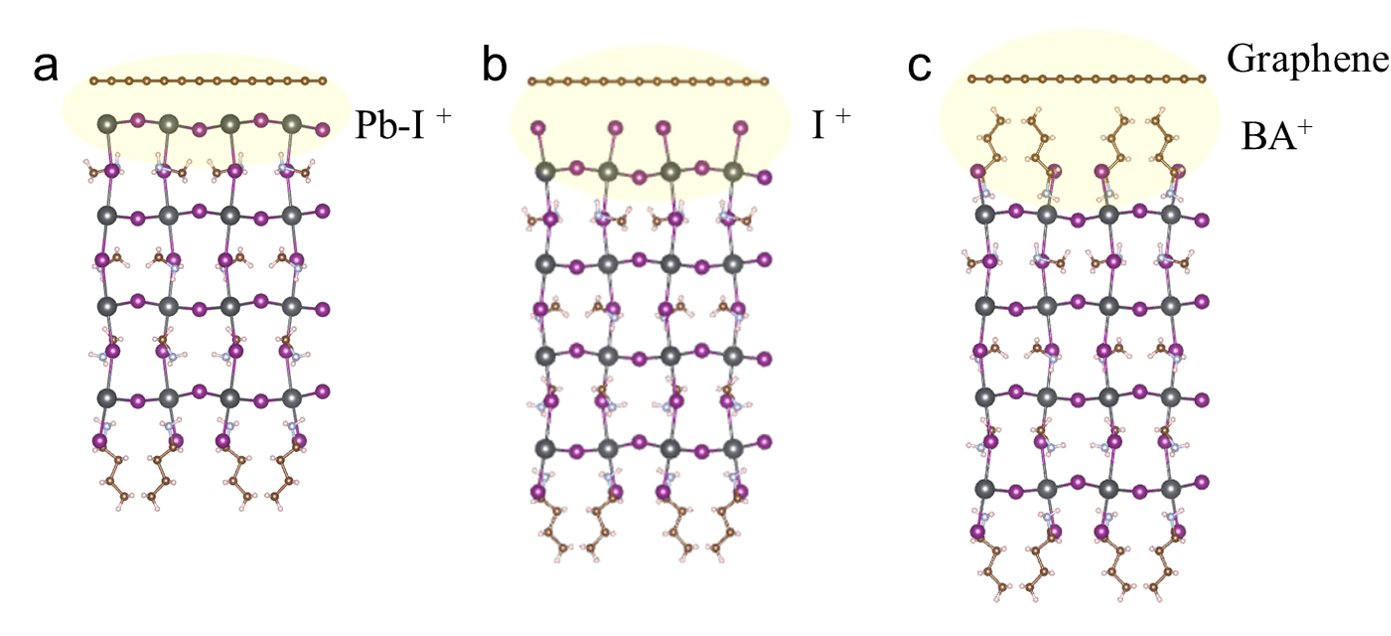} 
}
\caption{  \justifying 
When in contact with graphene, the termination of BMPI could be a) inorganic Pb-I facet, b) Idiode, and c) organic BA molecular. We calculated charge carrier density based on these three conditions and got the doping level of graphene in each case.
 } 
\label{fig:figs9}
\end{figure}

BMPI/Graphene heterostructures were built by stacking the BMPI slab on top of monolayer graphene along the y-axis, with terminations of BA molecular, Pb-I facet, and I atoms, respectively (Figure \ref{fig:figs9}). To make a commensurable heterostructure, artificial strain was required. It is known that small amounts of strain can have a sizable effect on the electronic properties of monolayer graphene \cite{si2016strain,choi2010effects}. Therefore, the strain was applied to the softer BMPI slabs \cite{guo2022atomistic,shao2022unlocking}. To limit the strain within 10 $\%$, a unit cell (Figure \ref{fig:figs8} ) consisting of a 1 $\times$ 2 $\times$ 1 of the perovskite slab and a 2 $\times$ 7 $\times$ 1 graphene was created. This unit cell required 3.1 $\%$ compression in the x direction and 4.8 $\%$ in the z direction of the BMPI slab.

\vspace{2.5mm}
{\bf Relaxation and Convergence}

we constructed three heterostructures as shown in Figure \ref{fig:figs9} and then fully relaxed the configuration. The convergence threshold for structure relaxation is 0.02 eV/Å. The interlayer distance between BMPI and graphene starts with 4 \AA and then fully relaxed to around 5 angstrom, which is similar to the reported value in perovskite heterostructure \cite{sun2020engineering}.

\vspace{2.5mm}
\noindent {\bf B.	Determination of Doping level}

\vspace{2.5mm}

{\bf by integration}
We calculated the electronic density. The output files in .cube format are saved for BMPI/graphene stack, pure graphene, and pure BMPI structure respectively, and then the differential charge density $\Delta \rho= \rho_{\rm graphene/BMPI} - \rho_{\rm graphene} - \rho_{\rm BMPI}$ can be obtained. We used Atomic Simulation Environment (ASE) for manipulating, visualizing, and analyzing atomistic simulations \cite{ase-paper}. The differential charge densities are spacially integrated to the number of electrons variation in graphene layer, which gives  $1.2 \times 10^{12}\; cm^{-2}$ for BA termination,  $1.2 \times 10^{13}\; cm^{-2}$ for Pb-I termination,  $3.9 \times 10^{13}\; cm^{-2}$ for I$^-$ termination. 

\vspace{2.5mm}
{\bf by Bader analysis}
Charge population can also be determined by the Bader charge analysis \cite{henkelman2006fast}. Taking the BMPI/graphene heterostructure with the termination of BA molecular as an example, there are 1256 electrons in one unit cell shown in Figure \ref{fig:figs8}, in which around 224 electrons in the graphene layer (The C pseudopotential indicates that two 1s electrons are fixed to the core, so each graphene contains 4 electrons, and there are 56 carbon atoms in graphene layer). The Bader charge analysis shows that 0.016 electrons are transferred from the BMPI layer to the graphene layer in forming the heterostructure with BA molecular as termination. As the unit cell in xz plane is  $8.52 \;\AA \times 17.22\;\AA$, on average graphene lost around $1.1 \times 10^{12}\; cm^{-2}$ electrons per square centimeters, which is close to the doping level from the experiment. 

\subsection{Photo-resistance spectrum}
\label{supinfoD}

\noindent {\bf A.	The calculation of the light-induced carrier density changes ($2.5 \times 10^{-5}$ ).}

According to Eq.1, the Resistance correction by Shubonikov-de-haas oscillation could be adapted to 

\begin{equation}
\delta R = A n^{-1} \cos\left( \frac{\pi h n}{2 e B} \right) \label{eqs1}
\end{equation}

, where $A = (\mu e)^{-1} D_T \exp \left(-\frac{\pi}{\mu B}\right)$, and the n is the carriers density. 

By taking the partial differential of the resistance with respect to carrier density, we get the optical-excitation-related resistance $\delta R^{p}$, which causes the carrier density changes.

\begin{equation}
\delta R^{p}  \approx -  A n^{-1}  \frac{ \pi h}{2 e B}\sin \left( \frac{\pi h n}{2 e B} \right) \delta n \label{eqs2}
\end{equation}
Considering the high carriers density, $ \frac{\pi h n}{2 e B} \simeq 200 \gg 1$, one can neglect the term coming from the derivative of $n^{-1}$ in Eq.~(\ref{eqs1}).

The light-induced resistance changes could be expressed by 

\begin{equation}
\sin \Delta \phi \frac{Amp(\delta R^{p})}{Amp(\delta R)} \approx \frac{\pi h n}{2 e B} \frac{\delta n}{n} \label{eqs3}
\end{equation}

here we only take the amplitude of two resistances and the sine component of the phase shift. From Figure 5b, we can get that the amplitude of photo-induced resistance $R^{p}$ is about 0.5 \% smaller than that of the resistance from SdH $\delta R$, and the phase shift is around 0.7 $\pi$. So, the light-induced carrier density changes $\frac{\delta n}{n} \approx 2.5 \times 10 ^{-5}$.

\vspace{5mm}
\noindent {\bf B.	The calculation of carrier temperature rising (0.3 K) by photo-resistance}

We deform Eq.1 into the temperature-related resistance variation. 
\begin{equation}
\delta R = B D_T \label{eqs4}
\end{equation}
, where 
\begin{equation}
B = (n \mu e)^{-1} \exp \left(-\frac{\pi}{\mu B}\right) \cos\left( \frac{\pi h n}{2 e B} \right) , 
\end{equation}
\begin{equation}
D_T = \frac{\Gamma}{\sinh \Gamma}= \frac{2 \pi^2 k_b T/\hbar \omega _c}{\sinh(2 \pi^2 k_b T/\hbar \omega _c)}
\end{equation}

By taking the partial differential of $\delta R$ versus Temperature (T), we get the light-induced temperature changes.
 
\begin{equation}
\delta R^{p} = B \delta D_T  \delta T \approx - \frac {1}{3}(\frac {2 \pi^{2} k_b}{\hbar \omega_c})^{2} T \delta  T \label{eqs4}
\end{equation}

The amplitude of temperature-induced resistance changes 

\begin{equation}
\cos \Delta \phi \frac{Amp(\delta R^{p})}{Amp(\delta R)} \approx  - \frac {1}{3}(\frac {2 \pi^{2} k_b}{\hbar \omega_c})^{2} T^{2} \frac {\delta  T}{T}
\end{equation}

The temperature-related resistance changes are mainly shown on the cosine component of the amplitude ratio. Taking the $Amp(\delta R^{p})/Amp(\delta R) = 0.005$ and $\phi = 0.7 \pi$ from the experiment, one can get $\frac {\delta  T}{T} = 0.2$. As the temperature T = 1.6 K in this measurement, illumination caused a temperature rise of around 0.32 K.

\vspace{5mm}
\noindent {\bf C.		The calculation of heating by Wiedemann-Franz law}

Because the sample is measured under low temperature (around 1.6 K) and away from the charge-neutrality point of graphene, the electronic thermal transport in our sample system can be described by Wiedemann-Franz law: 

\begin{equation}
k = L T / \rho ,
\end{equation}
where $\rho \approx 500 \Omega$ is the resistance per square under zero field and zero gate voltage, the L is Lorenz number and reported in graphene as $1.32 \times (\pi ^{2}/3)(k_b / e)^{2}$  ($\approx 3.22 \times 10^{-8} \; \Omega W K^{-2}$), $k_b$ is Boltzmann constant and $e$ is the elementary charge \cite{fong2013measurement}. we can get  thermal conductivity $k = 1.03 \times 10^{-10} W K^{-1}$ . The k is from the heat flux in a square piece of sample divided by the temperature variation in this square (Q = $-\alpha k \Delta T$). The heat flux basically comes from the laser excitation in our case, which is approximately equal to photon flux. 
In the photo-resistance measurement, the excitation power on the sample is around 40 $W/m^{2}$, and the thickness of the BMPI layer is around 14 nm (4 layers and assuming 15 $\%$ of light is absorbed \cite{leng2018molecularly}), so we can estimate the photon flux in the sample around 6 $W m^{-2}$. The $\alpha $ is a geometric factor related to the temperature non-uniform distribution between two electrodes (higher temperature in the middle and lower temperature near the leads). We assume a parabolic distribution of electron temperature, $\delta T (x) = 1/2 (1 - x^{2})$, and $\alpha ^{-1} = \int_{-l}^l \delta T (x) dx / \int_{-l}^l dx $ ( $2l$ is the distance between two contacts), which gives $\alpha = 3$. 
We take a square of $4 \times 4 \;\mu m^{2}$ (from the geometry of the bottom graphene connection) to calculate the general temperature change between two bottom electrodes. So $\Delta T (K) \approx 0.31\; K$

\noindent {\bf D.		The calculation of carriers' lifetime}
Under illumination, the variation of charge density $\delta n$ is proportional to the photon flux ($Q_{ph}$) and carriers' lifetime ($\tau$) with the formula   
\begin{equation}
\delta n = \frac{ Q_{ph}  }{h v} \times \tau
\end{equation}

$hv$ represents the energy of one photon. As $\delta n / n = 2.5 \times 10^{-5}$ and carrier density in our system is around  $ 2.8 \times 10^{-13} cm ^{-2}$, we can get the lifetime  $\tau \approx 400 \;ns$

\begin{figure} [h!]
\centerline{
\includegraphics[clip=true,width=16cm]{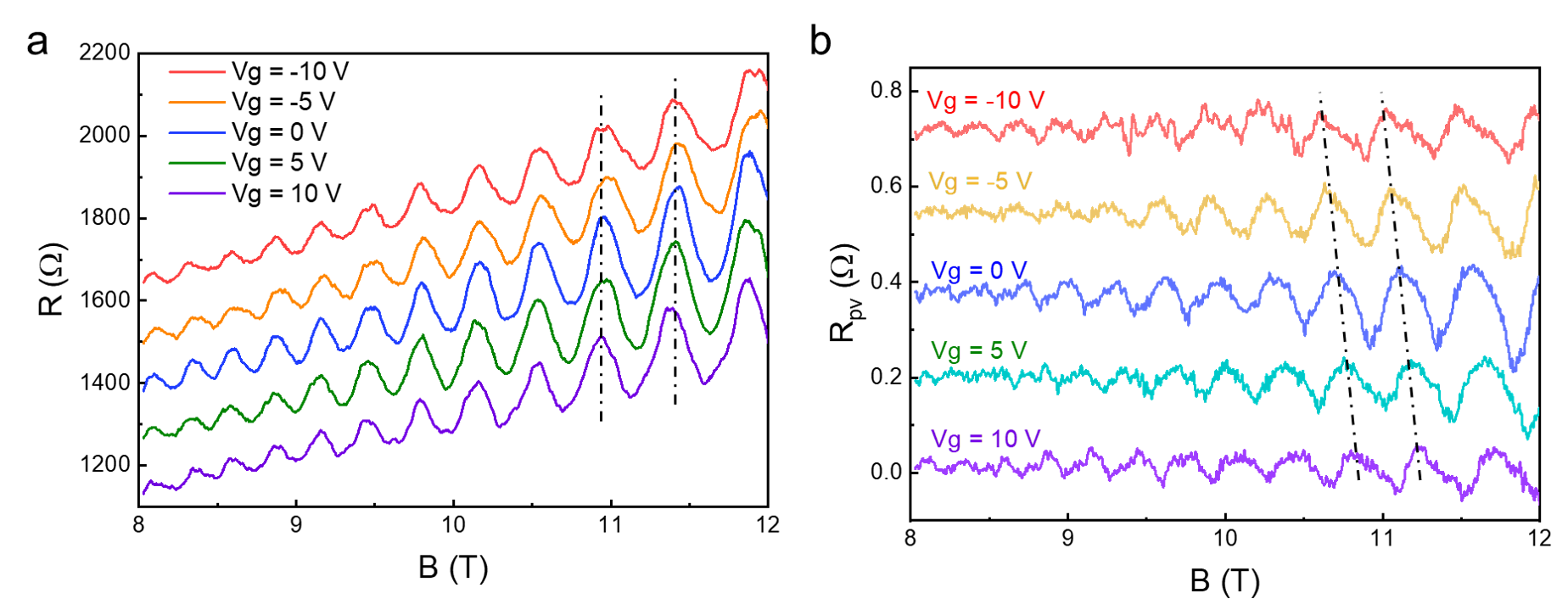} 
}
\caption{  \justifying 
a) The magneto-resistance and b) the photo-induced changes in magneto-resistance under several electric fields. The photo-induced resistance shift and the shape of the oscillation peaks become serrated while the oscillation features on the magneto-resistance have no obvious changes.
 } 
\label{fig:figs11}
\end{figure}

\end{document}